\documentclass[]{IEEEtran}
\usepackage{amsfonts}
\usepackage{amssymb}
\usepackage{graphicx}
\usepackage[dvipsnames]{xcolor}
\usepackage[caption=false,font=footnotesize]{subfig}
\usepackage{booktabs}
\usepackage{tabu}
\usepackage{paralist}
\usepackage{microtype}
\usepackage{float}
\usepackage{graphicx}

\usepackage{hyperref}
\usepackage[all=normal,paragraphs=tight,bibbreaks=tight,floats=tight]{savetrees}

\renewcommand{\paragraph}[1]{\medskip\noindent{\bf #1}}

\newcommand{\ignore}[1]{}

\newcounter{inlinecounter}

\begin{document}

\title{Your PIN Sounds Good! \\
 On The Feasibility of PIN Inference Through Audio Leakage}

\author{\IEEEauthorblockN{Matteo Cardaioli\IEEEauthorrefmark{1}\IEEEauthorrefmark{3},
Mauro Conti\IEEEauthorrefmark{1}\IEEEauthorrefmark{4}, 
Kiran Balagani\IEEEauthorrefmark{2}, and
Paolo Gasti\IEEEauthorrefmark{2}}

\IEEEauthorblockA{\IEEEauthorrefmark{1}University of Padua}
\IEEEauthorblockA{\IEEEauthorrefmark{2}New York Institute of Technology}
\IEEEauthorblockA{\IEEEauthorrefmark{3}GFT Italy}
\IEEEauthorblockA{\IEEEauthorrefmark{4}University of Washington, Seattle}}


\maketitle

\begin{abstract}

Personal Identification Numbers (PIN) are widely used as authentication method for systems such as Automated Teller Machines (ATMs) and Point of Sale (PoS). Input devices (PIN pads) usually give the user a feedback sound when a key is pressed. In this paper, we propose an attack based on the extraction of inter-keystroke timing from the feedback sound when users type their PINs. Our attack is able to reach an accuracy of 98\% with a mean error of 0.13 +/-6.66 milliseconds. We demonstrate that inter-keystroke timing significantly improves the guessing probability of certain subsets of PINs. We believe this represents a security problem that has to be taken into account for secure PIN generation. Furthermore, we identified several attack scenarios where the adversary can exploit inter-keystroke timing and additional information about the user or the PIN, such as typing behavior. Our results show that combining the inter-keystroke timing with other information drastically reduces attempts to guess a PIN, outperforming random guessing. With our attack, we are able to guess 72\% of the 4-digit PINs within 3 attempts. We believe this poses a serious security problem for systems that use PIN-based authentication.
	
\end{abstract}

\section{Introduction}
The use of authentication systems via Personal Identification Number (PIN) dates back to the mid-sixties \cite{batiz2011development}. The first devices to make use of this type of authentication were automatic dispensers and control systems at petrol stations, while the first applications in the banking sector came in 1967 with the cash machines \cite{bonneau2012birthday}. PINs have found great use over the years thanks to their suitability for for resource-constrained environments (i.e. numeric keypads instead of full keyboards) \cite{wang2017understanding}. Nowadays many people interact with devices that require authentication through PIN in different scenarios: Automated Teller Machines (ATMs), door-locking systems, Point of Sale (PoS), automated fuel dispensers, vending machines and mobile phones. For what concern financial services, basic security principles for PINs and for PIN entry devices (PIN pads) are specified in ISO 9564-1 \cite{iso9564}. Since PINs could suffer of shoulder surfing attacks \cite{roth2004pin,kumar2007reducing,kwon2015analysis}, to mitigate that, guidelines \cite{iso9564} indicates that values of entered PIN shall not be displayed in plain text or be disclosed by audible feedback. In order to reach a compromise between usability and security, the common behavior of interfaces/devices is to mask video feedback \cite{silktv} and to provide always the same feedback sound for all keys. In particular, that assumes that recording sound feedback does not give any additional information about the PIN. 

In this paper we demonstrate it is possible to obtain information about the PIN from the feedback sound by deducing the inter-keystroke timing. Unlike a direct attack (input-based), in which the adversary observes the PIN pad while the user enters the code, the sound recording (output-based) does not require a direct view of the input device. This gives to the adversary a significant advantage in terms of exposure: making the adversary difficult to identify. For example, the adversary can record the feedback sound of an ATM maintaining te courtesy distance. No special devices are required for sound recording, but a smartphone or a voice recorder may be sufficient. Leveraging this attack, the adversary is able to reduce significantly the attempts to guess the victim's PIN with respect to a random guessing. 
In this paper, we extend the inter-keystroke timing algorithm presented in \cite{pilot} with several additional information that can drastically reduce the attempts to guess the PIN:

\begin{enumerate}

\item {Knowledge about typing behavior: it is behavioral feature that takes into account how a user interacts with the PIN pad (e.g., user types the PIN always with the same finger). This information can be easily acquired by observing the victim in different contexts and not necessarily when entering the PIN in a ATM, but in any situation in which he interacts with a PIN pad (e.g., door-lockers, automatic fuel pumps, laptop numeric keyboard).}	

\item {Knowledge of one digit of the code: the adversary has managed to spy the victim during the PIN entry and has come into possession of the value and position of a number in the code sequence. This information can be obtained by observing the victim while entering the PIN. Depending on the environment, input devices do not usually offer an adequate coverage, resulting more vulnerable to input-based shoulder surfing attack. Furthermore, PoS are often used in crowded place where the user tends to be distracted (e.g., supermarkets) and generally does not pay particular attention during typing.}

\item	{Knowledge of the values composing the PIN, but not their order: one of the techniques applicable in this scenario is the thermal attack. Once the victim has completed his operations on ATM or PoS the adversary acquires a thermal image of the PIN pad. Subsequently, through the analysis of the heat map, the adversary can determine the keys pressed. Thermal camera are on the market at low cost (i.e., FLIR ONE Gen 3 camera costs less than \$ 200 \cite{thermalCamera}) and can be also used as smartphone devices.}

\end{enumerate}

\paragraph{Contributions.}
\begin{itemize}
\item We demonstrate that it is possible to retrieve fairly accurate inter-keystroke timing information from recorded-audio with common devices. We are able to correctly detect 98\% of feedback sounds with an error of 0.13 +/- 6.66ms. Furthermore, 75\% of inter-keystroke timings extracted by the software had absolute errors under 7.66ms.

\item We study and analyze how the interaction behavior of the user with the PIN pad affects the guessing of a PIN. We show that users who type PINs with just one finger are more vulnerable to an attack that exploits interkeystroke timing. We also show how the combination of inter-keystroke timing and single finger typist knowledge leads to a 34-fold improvement over random guessing to guess a PIN within the first 5 attempts.

\item We propose and study, for the first time in the literature, the application of inter-keystroke timing to different types of attacks. We show that inter-keystroke timing significantly improves performance for each attack. For example, applying inter-keystroke timing to Thermal Attack we are able to guess 15\% of the PINs at the first attempt, reaching a fourfold improvement in performance. We also show how the combination of multiple attacks can dramatically reduce attempts to guess the PIN. We are able to guess 72\% of the PINs within the first 3 attempts.

\item In this work, we show how the inter-keystroke timing is a factor that influences the probability distribution of PINs, making some of them much more likely than others. This poses a serious security problem for PIN-based authentication devices.
\end{itemize}

\paragraph{Organization.} 
Section~\ref{sec:related_work} overviews
state-of-the-art in password guessing based on acoustic attacks and on non-acoustic side channel attacks. Section~\ref{sec:system_adversary} presents our system and the adversary model. Section~\ref{sec:soundanalysis} describes an algorithm to extract inter-keystroke timing from an audio signal.
In Section~\ref{sec:pin_guessing} we report our results in PIN guessing, merging different types of knowledge about the PIN. In Section~\ref{sec:pin_dist} we analyse how the different types of knowledge affect the randomness of PINs. The paper concludes with the summary and future work
directions in Section~\ref{sec:conclusion}.

\section{Related Work}\label{sec:related_work}
Non-acoustic side channel-attacks represent a type of approach to password inference. Vuagnoux and Pasini \cite {vuagnoux2009compromising} showed how it is possible to recover keystrokes by analysing the electromagnetic waves emitted by the electrical components of the keyboard. Marquardt et al. \cite {marquardt2011sp} collected with an accelerometer the vibrations caused by the surface under the keyboard while typing to infer the pressed keys. Other attacks focused on motion detection using embedded sensors on mobile devices. Sarkisyan et al.\cite {sarkisyan2015wristsnoop} and Wang et al. \cite {wang2016friend} inferred smartphone PINs through the analysis of smartwatch motion sensors such as accelerometers and magnetometers. 
All the attacks presented so far in this section, implys that the adversary must acquire the information when the user is entering the password. However, there are attacks that allow to obtain information about the password seconds after it has been typed.  These non-acoustic attacks are based on thermal heat transfer and thermal emanation. They were introduced for the first time by Zalewski \cite {zalewski2005cracking}. Every time a user presses a key the heat of the finger is transferred to the keyboard and can be recorded with a thermal camera. Depending on the material of the keyboard, thermal residues have a different dissipation speed \cite {mowery2011heat}, changing the time window in which the attack can be effective. Abdelrahman et al. \cite {abdelrahman2017stay} evaluated how different PINs and patterns on mobile device can influence thermal attack performance. Kaczmarek et al. \cite {kaczmarek2018thermanator} demonstrate how a thermal attack can recover precise information about a password within 30 seconds it was entered and partial information within 60 seconds. 

Acoustic attacks represent another type of approach to password inference. The first to study vulnerability on keyboard, relying on the fact that each key emits a characteristic sound,  were Asonov and Agrawal \cite{asonov2004keyboard}. They trained a neural network on features extracted by the recorded audio signal to classify each key pressed. Subsequent work showed the efficacy of sound emanation.  Berger et al. \cite {berger2006dictionary} combined keyboard acoustic emanation with a dictionary attack to reconstruct single words, while Halevi and Saxena \cite {halevi2012closer} studied keyboard acoustic emanations to eavesdrop over random password. Unfortunately the recognition of the keys pressed by identifying the particular sound that each of them produces cannot be taken into consideration in our case. 

Another kind of acoustic attacks is based on time difference of arrivals (TDoA) \cite{zhu2014,wang2014ubiquitous,liu2015snooping}. They rely on multiple microphones recording where TDoA is used to triangulate the position of the pressed key. This type of attacks, although allow to obtain a good accuracy, are difficult to implement in real situations. Involved hardware and environmental preparation could limit their applicability. Song et al. \cite {song2001timing} presented an attack based on latency between subsequent key pressed, showing that information about inter-keystroke timing can be used to substantially narrow the search space. A similar approach was used by Balagani et al. \cite {silktv} who deduced inter-keystroke timing from the time of appearance on the video of the masking symbols when a user is typing a password.

The closest work to this paper is \cite {pilot}, in which is demonstrated that precise inter-keystroke timing information drastically reduce attempts to guess a PIN. Moreover, deducing information of inter-keystroke timing from videos can reduce up to 26 times the number of attempts to guess a PIN compared to random guessing. The limitation of \cite {pilot} is the use of cameras to record the ATM screen while a user was entering the PIN. This reduces the number of devices that can be attacked to those that actually show clear graphic feedback, making other PIN based devices very difficult to attack  (i.e. PoS has very small and dim screens). 

In this paper we combine inter-keystroke timing deduced from sound recording with other information deduced from other plausible attacks in a real context. This setting can be applied to a multitude of devices that can be attacked, drastically reducing the attempts to guess a PIN compared to \cite {pilot}. To our knowledge there is no other work that use the strategy presented in this paper, making our attack scenario unexplored. The simple setting of this attack makes it easy to apply in real contexts, posing immediate and serious implications in systems like ATM or PoS.

\section{System and Adversary Model}\label{sec:system_adversary}
In this paper, we modeled an authentication system via Personal Identification Number (PIN) that simulates the log-in process of an Automated Teller Machine (ATM) through interaction with a PIN pad (input device). For each key pressed when entering the secret code, the user immediately receives a feedback sound (output). Duration and frequency of the sound do not depend on the key pressed and are always constant, following ISO 9564-1 standard \cite {iso9564}. The goal of the adversary is to learn the victim’s PIN. The adversary can record the output without having visibility of the input device or the screen. Since a direct view of the PIN pad is not necessary, information can also be obtained even if the victim covers it with his hand while entering the code. Furthermore, in order to record the feedback sound, the adversary can be located at a distance of a couple of meters, also in a noisy enviroment. This attack setting can be referred to a multitude of real situations in which the adversary uses a common recording device (such as a smartphone): 1) the user is withdrawing from an ATM using bank card and the adversary is waiting in line for his turn; 2) a customer is paying in a shop via PoS and the adversary is near the device; 3) the victim is opening a door with a electronic PIN lock. In this case the adversary could have a hidden recording device and turn his back to the victim. Furthermore, the adversary may not even be physically close to the victim. For example: 1) the adversary has access to the audio of cameras present in the proximity of a terminal that requires authentication via PIN; 2) the adversary hides a recording device, which can also be accessed remotely, near an ATM or any device provided with a PIN authentication method.
We have also identified several information leakage about the user or the PIN the adversary can exploit in the attack:

\begin{enumerate}
\item	Knowledge about typing behavior.
\item	Knowledge of one digit of the code.
\item	Knowledge of the values composing the PIN, but not their order. 
\item	Knowledge about typing behavior and knowledge of one digit of the code: in this scenario the adversary is in possession of both the information described in points 1 and 2.
\item	Knowledge about typing behavior and knowledge of the values composing the PIN, but not their order: in this scenario, the adversary is in possession of both the information described in points 1 and 3.
\item	Knowledge of the values composing the PIN, but not their order and knowledge of one digit of the code: in this scenario, the adversary is in possession of both the information described in points 2 and 3.
\item	Knowledge about typing behavior, knowledge of one digit of the code and knowledge of the values composing the PIN, but not their order: in this scenario the adversary is in possession of all the information described in points 1, 2 and 3.
\end{enumerate}
The full attack proceeds as shown in Figure \ref{fig:STEP}:
\begin{figure}[h!]
	\centering
		\includegraphics[trim = 1.9in 2in 0.1in 2in, clip, scale=0.57]{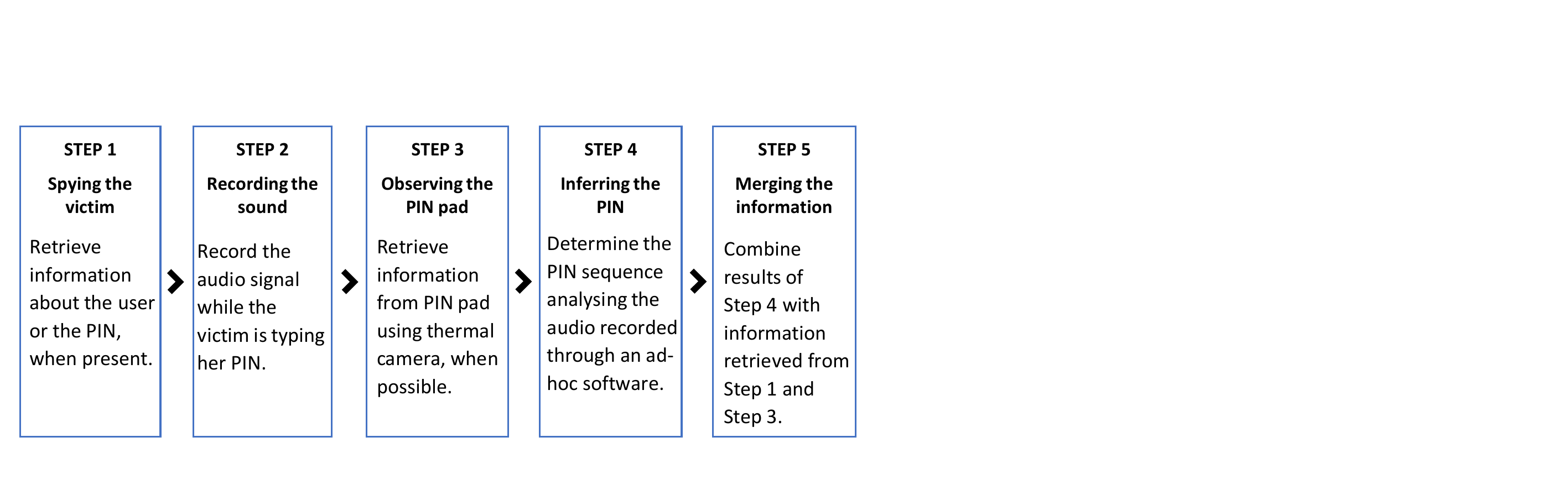}
	\caption{Attack sequence.}
	\label{fig:STEP}
\end{figure}

\section{Key-stroke detection through sound analysis}\label{sec:soundanalysis}
We used the same dataset presented in \cite {silktv} to infer inter-keystroke timing. Twentytwo participants were asked to type 4-digit PINS on a simulated ATM. Nineteen of them completed the full task consisting of three data collection sessions while 3 completed only one session. For each session participants had to digit a total of 180 PINs following this procedure:
\begin{enumerate}
\item The participant memorizes a 4-digit PIN that appears on the screen for 10 seconds.
\item Once the PIN in point 1 disappears, the participant types it 4 times.
\item Points 1 and 2 are repeated 15 times.
\item The sequence described in points 1, 2 and 3 is repeated 3 times using the same PINs. As a result, each PIN is typed 12 times per session.
\end{enumerate} 
Whenever a participant pressed a key, the ATM simulator emitted a feedback sound and logged the relative timestamp with millisecond precision. Sessions were recorded in a noisy environment (SNR -15 dB) with a Sony FDR-AX53 camera located 1.5 m away from the PIN pad.  Audio signal was acquired with a sampling frequency of 48000 Hz.  

\begin{figure}[h!]
	\centering
		\includegraphics[trim = 0.1in 2.5in 0.1in 2.5in, clip, scale=0.4]{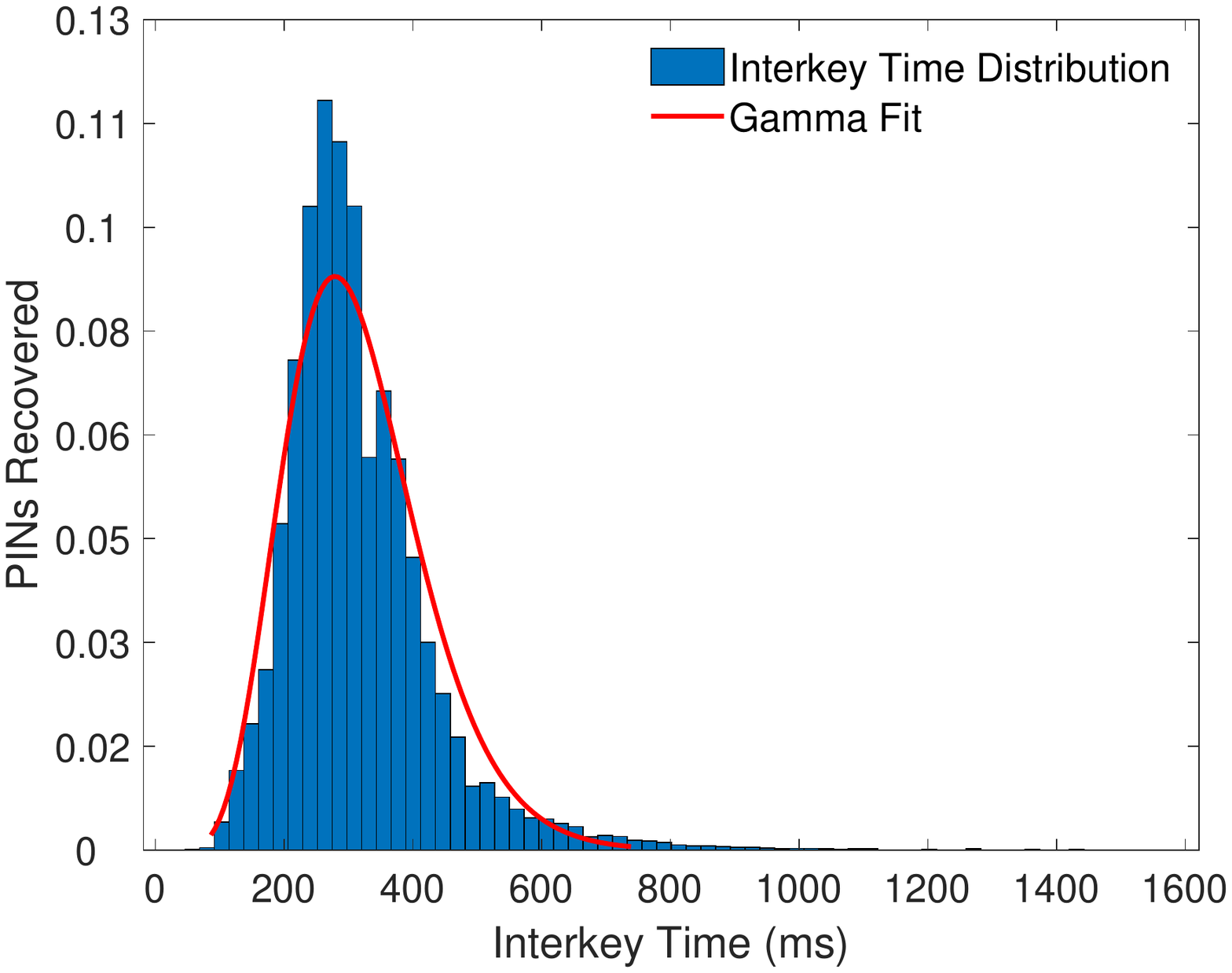}
	\caption{Logged inter-keystroke timing distribution modelled as gamma distribution.}
	\label{fig:interkey_gamma}
\end{figure}

\begin{figure}[h!]
	\centering
		\includegraphics[trim = 0.1in 2.5in 0.1in 2.5in, clip, scale=0.4]{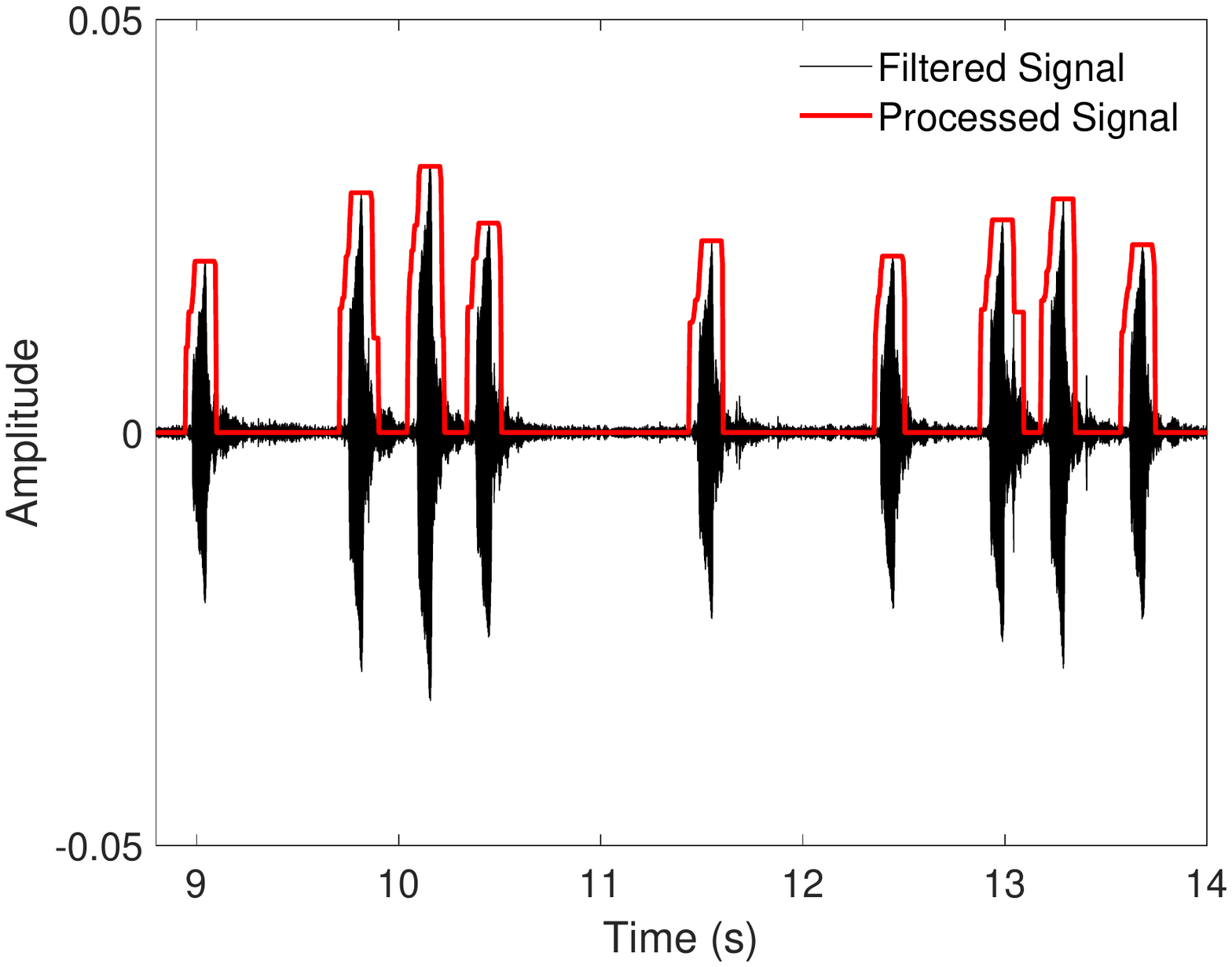}
	\caption{Comparison between filtered sound signal and processed signal.}
	\label{fig:processed_sound_signal}
\end{figure}

\begin{figure}[h!]
	\centering
		\includegraphics[trim = 0.1in 2.5in 0.1in 2.5in, clip, scale=0.4]{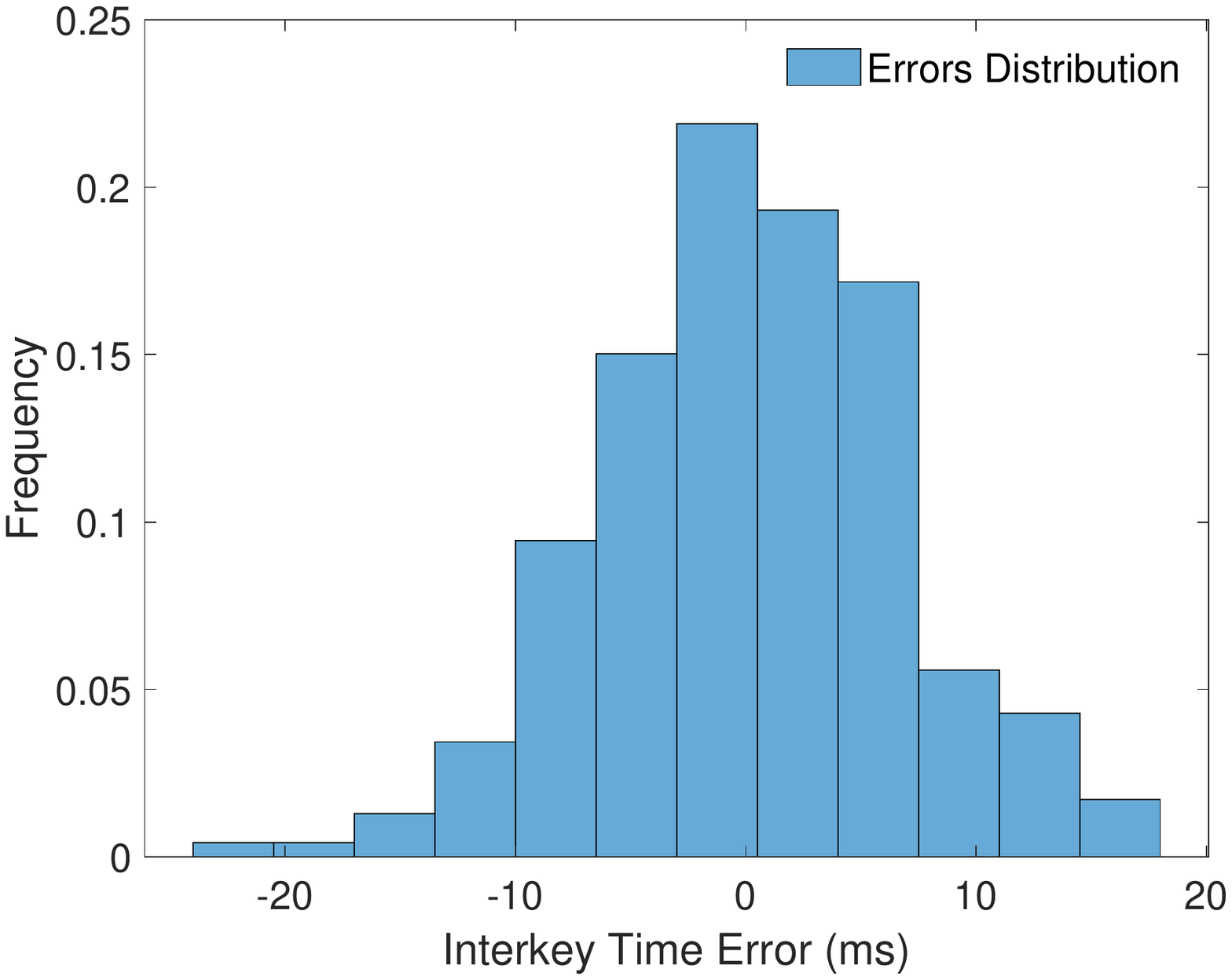}
	\caption{Errors distribution of our sound extraction algorithm.}
	\label{fig:extimation_error_dist}
\end{figure}

We developed a software to analyze the audio recorded and to detect the instant in which a key was pressed. The signal was linearly normalized in amplitude in an interval ranging from -1 to +1. We applied a 16 order Butterworth band-pass filter \cite{butterworth1930theory} centred at 5600 Hz to the recorded signal in order to isolate the characteristic frequency window of the feedback sound. We processed the filtered signal setting all the values below a threshold of 0.01 in amplitude, to remove the residual noise to 0. We then calculated the maximum across nearby values in a sliding window of length 4800 samples (4799 overlapping samples), corresponding to 100 milliseconds (one-tenth the sampling frequency). To determine the length of the window we evaluated the distance between consecutive timestamps logged by the ATM simulator. These differences follow the probability distribution showed in Figure \ref{fig:interkey_gamma}, which can be modelled with a gamma function. We set a threshold of 0.001 to discard outliers that could cause the occurrence of false positive or negative during sound processing. From this we deduced that 99.9\% of the differences between timestamps were higher than 100ms (4800 samples). In Figure \ref{fig:processed_sound_signal} we present the result of the process described above. We extracted the timestamps of the peaks of the processed signal and compared to the ground truth. All the estimated values that differed by more than 25ms from the real value were classified as errors. We than performed the Anderson-Darling test \cite{anderson1952asymptotic} that confirmed the whiteness of the residuals. Our experiments show that our algorithm leads to fairly accurate inter-keystroke timing information. It is able to correctly detect 98\% of feedback sound with an error of 0.13 +/- 6.66ms. Furthermore, 75\% of inter-keystroke timings extracted by the software had absolute errors under 7.66ms (against 10ms of \cite {silktv}), and 97\% had errors under 13.06ms (against 20ms of \cite {silktv}).Figure \ref{fig:extimation_error_dist} shows the errors distribution.

\section{PIN guessing}\label{sec:pin_guessing}
In this section we first introduce and describe an algorithm for PIN-guessing\cite {pilot}, based on the relationship between inter-keystroke timings and Euclidean Distance of consecutive keys. Than we present our results applying this algorithm to the scenarios introduced in Section \ref {sec:system_adversary} related to additional information about the user or PIN. 

\subsection{Base Attack}\label{sec:base_attack}
Base Attack \cite {pilot} is an algorithm that ranks PINs from the estimated distances between subsequent keys. Keys are considered as nodes of a multigraph, while Euclidean Distances between subsequent keys are considered as weighted edges. The possible distances are: zero distance (0), distance one (1), distance two (2), distance three (3), diagonal distance one (d1), diagonal distance two (d2); short diagonal distance (sd) and long diagonal distance (ld). Only one distances triplet can be associated to a 4-digit PIN (e.g., to PIN 5566 is associable the only triplet 0,1,0), while more PINs can be associated to a triplet. Base Attack removes all non-compliant path to the triplet received as input, returning as output all the associated PINs to the input triplet (e.g., for the input 0,3,0 the algorithm returns PINs 2200 and 0022). Base Attack significantly reduce the attempts to guess a PIN, reaching a 26-fold improvement in peformance compared to random guessing. Moreover, \cite{pilot} demonstrates a correlation between Euclidean Distance and inter-keystroke timings.

\subsection{Knowledge of typing behavior}\label{sec:typing_behavior}
Base Attack demonstrate a correlation between inter-keystroke timings and Euclidean Distance. However, it is reasonable to expect that this correlation is linked to the way users type the code. Suppose that the user has to enter a PIN containing the keys 1 and 0 in sequence (e.g., 1099). Following the configuration of our PIN pad, the Euclidean Distance between keys 1 and 0 is the largest one and therefore we expect the inter-keystroke timing to have relatively high values. This is true if the user enters the first digit with one finger and the next one always with the same, actually covering the distance between one key and the other. It would no longer be the case if, for example, he would type the first digit with the index finger of the right hand and the second with the index finger of the left hand. In this case the inter-keystroke timing is supposed to be significantly shorter. To demonstrate this, we first studied the behavior of the 22 participants and then evaluated the contribution of this information to PIN guessing. We analyzed the 61 videos recorded in the data collection phase and for each PIN entered by each user we noted the typing behaviors, dividing them into: single finger or other. We classified all the PINs entered always and only with the same finger as “single finger”. This includes the codes entered with the index finger only and those typed with the thumb only. “Single finger PIN” (SFP) represent 70\% of the total, of them, 92\% were typed using the index while 8\% are typed with the thumb. The class “other PIN” (OP) represents the 30\% of the total. This class can be divided in the following subclasses: 

\begin{itemize}
\item \textit{Two hands PIN}: PINs typed with at least a finger of the left hand and at least a finger of the right hand. They represents the 38\% of the OP class.
\item \textit{One hand PIN}: PINs typed with at least two finger of the same hand. They represents the 34\% of the OP class.
\item \textit{Unknown typing behavior PIN}: all the PINs that we were unable to classify with certainty due to the total or partial coverage of the PIN pad during video recordings. They represents the 28\% of the OP class.
\end {itemize}

\begin{figure}[h!]
	\centering
		\includegraphics[trim = 0.1in 2.5in 0.1in 2.5in, clip, scale=0.4]{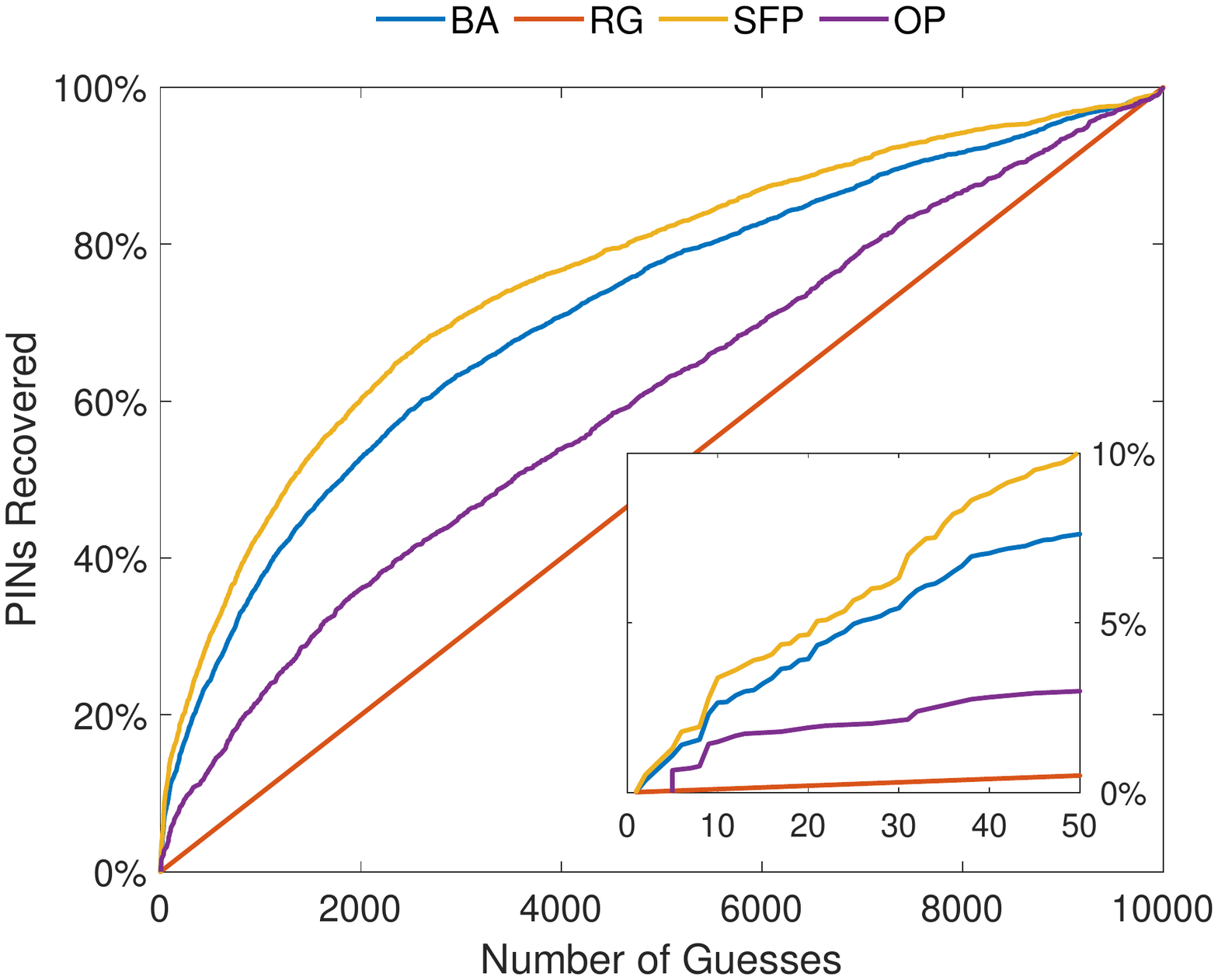}
	\caption{CDF showing the percentage of PINs recovered with a Base Attack (BA) compared to Single Finger PIN  (SFP), Other PIN  (OP) and the baseline (RG) .}
	\label{fig:finger}
\end{figure}

We found that 16 users on 22 (73\%) typed more PINs using a single finger (single finger typist - SFT), while the remaining 6 (27\%) typed more PINs using two hands or more than a finger of one hand (other typist - OT). We have also noticed that users have an attitude to enter PINs always in the same way. In particular, SFT typed 89\% of the time using just one finger and OT typed 77\% the time using 2 hands or more finger of a hand. 
We split our keystroke dataset in two sets. The first (training set) consists of 5195 PIN, typed by 11 participants. The second (testing set) consists of 5135 PIN, typed by a distinct set of 11 participants and filtered by SFP (3461 PINs) and OP (1674 PINs). 
Figure \ref{fig:finger} shows the effectiveness of knowing if a PIN is typed with only one finger compared to Base Attack, to OP and to random guessing. Our results show a significant difference between SFP and OP, while a more limited improvement between SFP and Base Attack (see FIgure \ref {fig:finger}). In particular, the percentage of PIN recovered within 5 attempts is doubled if the code is an STP instead of an OP. Moreover SFP shows a 34-fold improvement compared to random guessing to guess PINs within 5 attempts.

\subsection{Knowledge of one digit and its position in the code}\label{sec:VPK}
\begin{figure}[h!]
	\centering
		\includegraphics[trim = 0.1in 2.5in 0.1in 2.5in, clip, scale=0.4]{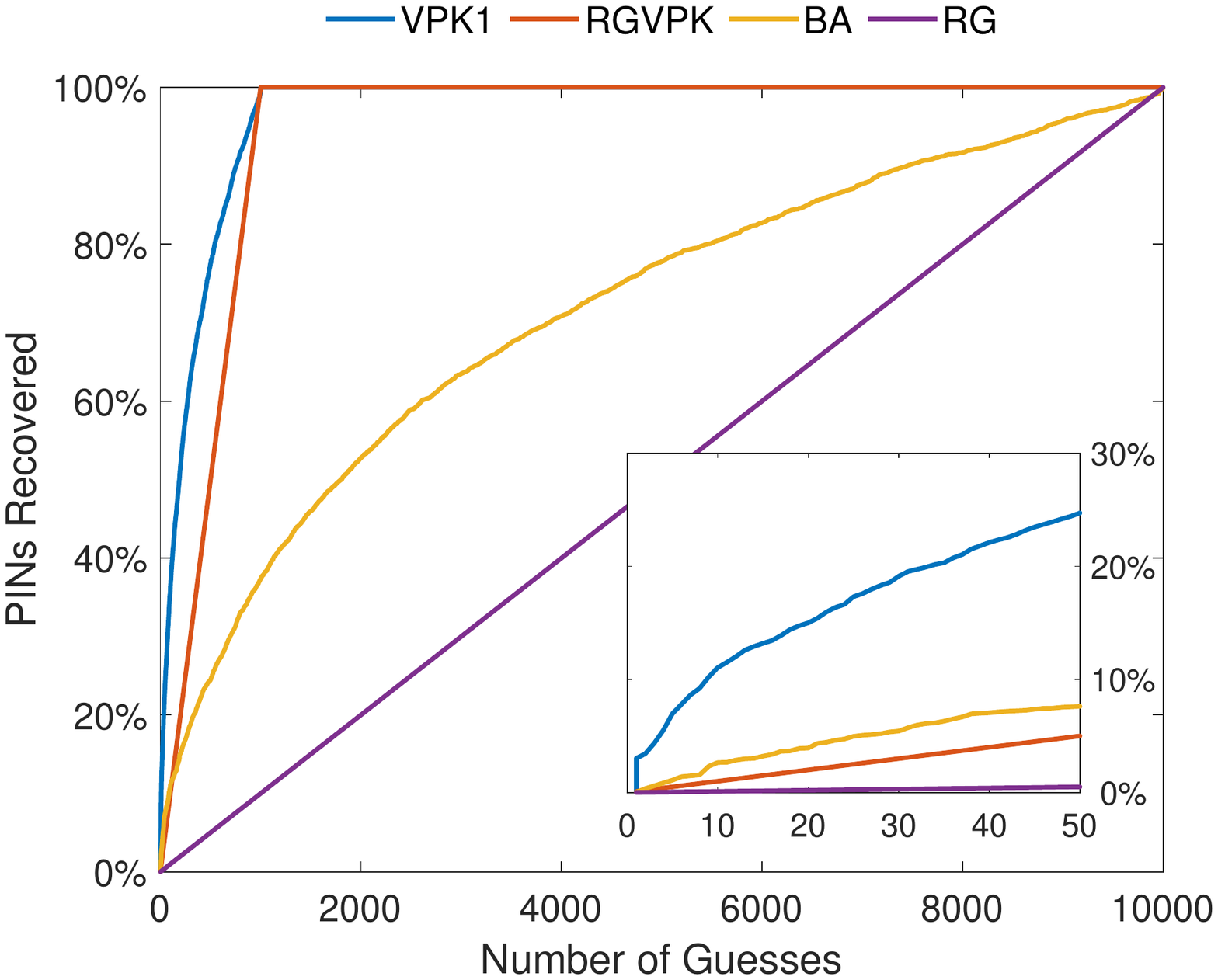}
	\caption{CDF showing the percentage of PINs recovered with the Value-Position Knowledge of the first digit (VPK1) , compared to random guessing value-position knowledge (RGVPK), Base Attack (BA) and Random Guessing (RG).}
	\label{fig:VPKvsRGVPK}
\end{figure}

We assume that the adversary has come into possession of the value and the position of one digit from the PIN (VPK- Value-Position Knowledge). We investigated how effective VPK is on real data, compared to RGVPK random guessing value-position knowledge. We split our keystroke dataset in two sets. The first (training set) consists of 5195 PIN, typed by 11 participants. The second (testing set) consists of 5135 PIN, typed by a distinct set of 11 participants. For each PIN in the testing set, we associated a list of triplets of distances sorted by their probability, like for the Base Attack. For each input triplets the algorithm returns a set of associated PINs. Since we are in possession of both the value and the position of a digit of the PIN, we filtered the output removing the sequences that were not compliant. For instance, given estimated distances 3, 0, and $\sqrt{2}$, the possible associated PINs are: 0007, 0009, 2224 and 2226. If we know that the first digit of the “real PIN” is 2, we can restrict the scope to 2224 and 2226.  Figure \ref {fig:VPKvsRGVPK} shows the effectiveness of this algorithm comparing VPK to RGVPK.

\begin{figure}[h!]
	\centering
		\includegraphics[trim = 0.1in 2.5in 0.1in 2.5in, clip, scale=0.4]{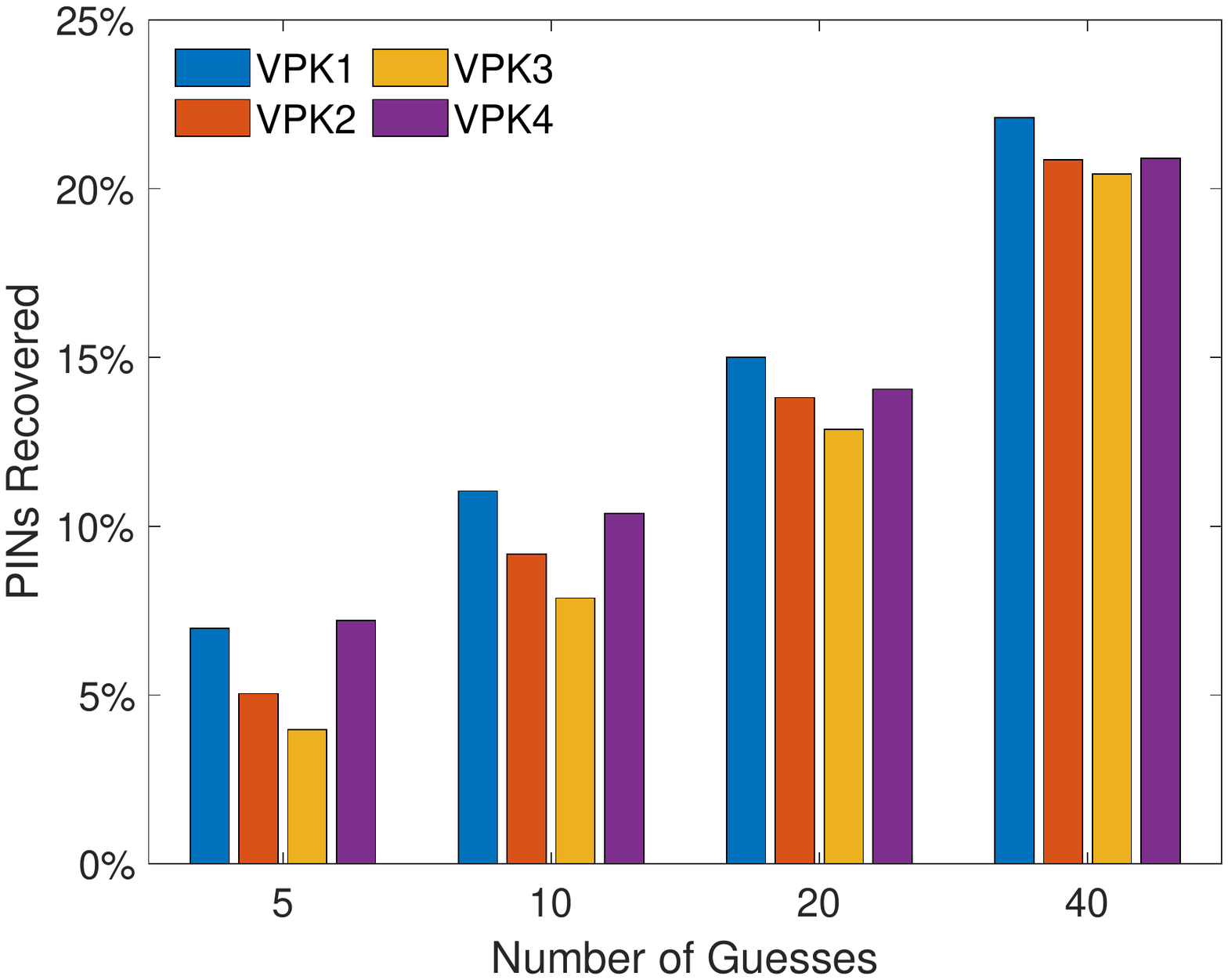}
	\caption {Percentage of PIN guessed knowing the first digit (VPK1), the second digit (VPK2), the third digit (VPK3) and the fourth digit (VPK4).}
	\label{fig:VPK1234_hist}
\end{figure}

The knowledge of a digit of a 4-digit PIN and its position, regardless of which is the first, second, third or fourth, reduces the number of PINs to be guessed from 10,000 to 1,000. This also applies theoretically to both the CDF of RGVPK and the CDF of VPK. In a real context, for the adversary, it is easier to get the value of the first digit of the PIN than the other 3. However, we wanted to investigate in our data set whether the knowledge of the position and of the respective digit was influential. We then performed 4 VPK attacks, varying the known position of the digit: VPK1 knowledge of the first digit, VPK2 knowledge of the second digit, VPK3 knowledge of the third digit and VPK4 knowledge of the fourth digit. Results are shown in Figure \ref {fig:VPK1234_hist}. 

We performed a series of $\chi^2$ tests to evaluate differences in the frequency of the PINs guessed for the 4 attacks. We found that:

\begin{itemize}
\item The $\chi^2$ statistic for VPK1 and VPK2 resulted significant at p$<$0.01 for the firsts 14 attempts.
\item The $\chi^2$ statistic for VPK1 and VPK3 resulted significant at p$<$0.01 for the firsts 31 attempts.
\item The $\chi^2$ statistic for VPK1 and VPK4 never resulted significant (p$>$0.01).
\item The $\chi^2$ statistic for VPK2 and VPK3 resulted significant at p$<$0.01 for the firsts 6 attempts.
\item The $\chi^2$ statistic for VPK2 and VPK4 resulted significant at p$<$0.01 for the firsts 8 attempts.
\item The $\chi^2$ statistic for VPK3 and VPK4 resulted significant at p$<$0.01 for the firsts 16 attempts.
\end{itemize}

Results show that, for our data, the knowledge of the position significantly influences the percentage of PINs guessed by the algorithm. In particular, knowing the first or the fourth digit gives more information than the other two digits, while the knowledge of the third digit generally gives a smaller contribution compared to others.

\subsection{Knowledge of the values composing the PIN, but not their order}\label{sec:thermal}
In this scenario, we assume that the adversary is aware of the keys pressed by the victim, but not of the temporal order in which they were pressed. This type of knowledge can be obtained by spreading powder on the PIN pad. Than, once the victim has finished her operations, the adversary can check the keys in which the powder has been removed. Another way to retrive the information of the values composing the PIN is a Thermal Attack (TA). As described in \cite {kaczmarek2018thermanator}, the adversary, using a thermal camera, records the heat footprint (heatmap) left by the victim on the PIN pad once she has finished her operation. In this work we want to investigate the contribution of this knowledge on real data, combining it with the Base Attack.
Suppose that the conditions allow the adversary to perform a TA on the PIN pad. He must therefore be in one of the following situations:

\begin{enumerate}
\item The heatmap shows only one heat zone, the victim then typed the same key 4 times (Class 1).  Suppose that the adversary deduces from the heatmap that the victim has typed key 5. The only PIN that can be associated in this context is 5555. In Class 1 the number of possible PINs is reduced to 1.

\item The heatmap shows 2 distinct heat zones, this means that the victim typed 2 different digits, repeating both twice or repeating one of them for 3 times (Class 2). Suppose that the adversary deduces from the heatmap that the victim has typed keys 0 and 2. Among the PINs that can be associated there are, for example, 0022, 0202, 0002, etc.. In Class 2 the number of possible PINs is equal to the permutations of 2 values (A and B) in 4 extractions excluding the AAAA and BBBB combinations that belong to Class1. The number of possible PINs is therefore reduced to 14 ($2^4-2$).

\item The heatmap shows 3 distinct heat zones, this means that the victim typed 3 different digits one of which was repeated twice (Class 3). Suppose that the adversary deduces from the heatmap that the victim has typed keys 0, 2 and 5. Among the PINs that can be associated there are, for example, 0255, 0225, 5502, etc.. In Class 3 the number of possible PINs is equal to the permutations in four extractions of the three digits deduced from the heatmap for the number of permutations of the repeated digit. The number of possible PINs is therefore reduced to 36 ($4\cdot3\cdot3$).

\item The heatmap shows 4 distinct heat zones, this means that the victim typed 4 different digits (Class 4). Suppose that the adversary deduces from the heatmap that the victim has typed keys 0, 2, 5 and 8. Among the PINs that can be associated there are, for example, 0258, 2580, 0852, etc.. In Class 4 the number of possible PINs is equal to the simple permutations of 4 elements. The number of possible PINs is therefore reduced to 24 ($4!$).
\end{enumerate}

\begin{figure}[h!]
	\centering
		\includegraphics[trim = 0.1in 2.5in 0.1in 2.5in, clip, scale=0.4]{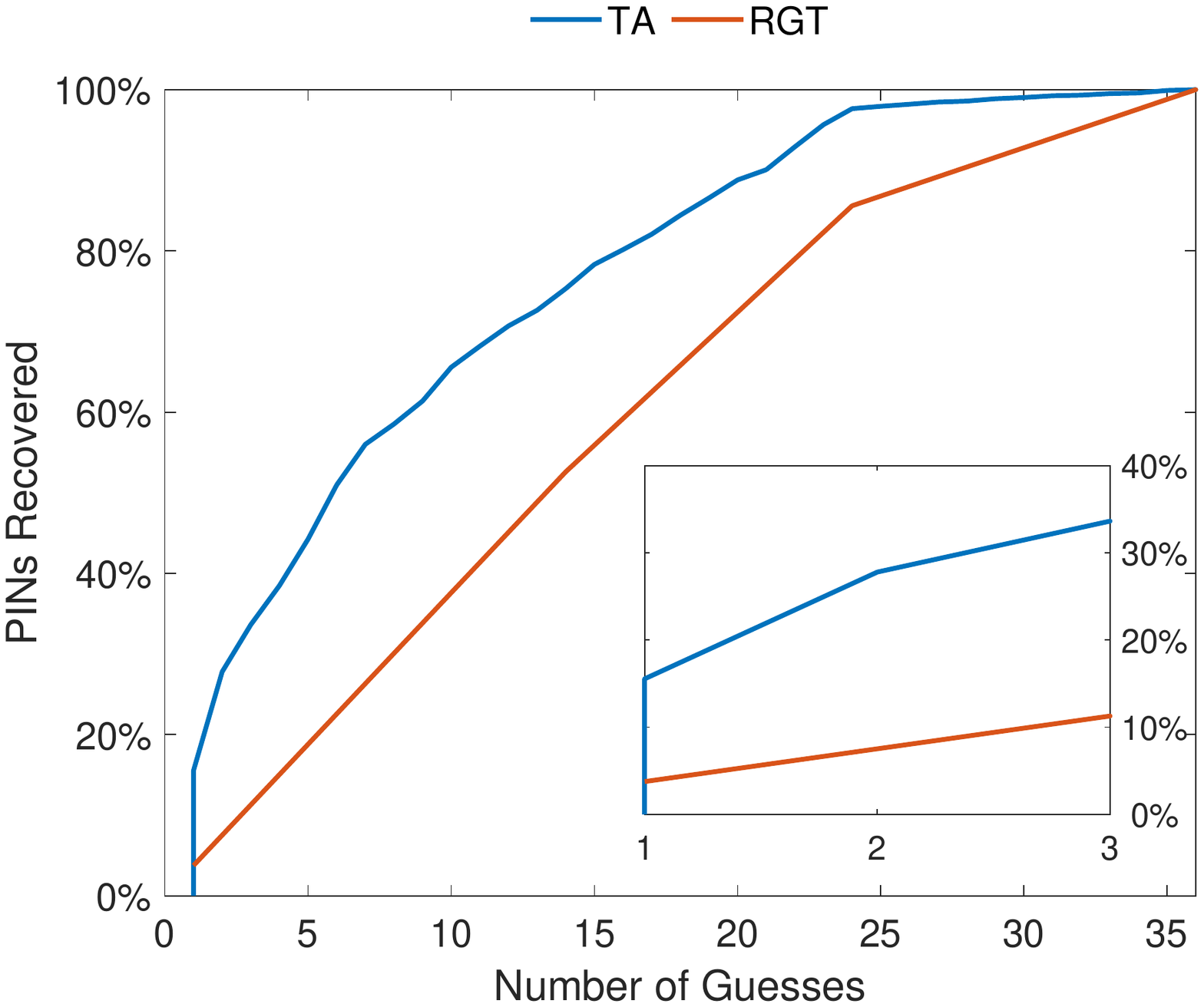}
	\caption{CDF showing the percentage of PINs recovered with a Thermal Attack (TA) compared to Random Guess Thermal (RGT).}
	\label{fig:thermal}
\end{figure}

We applied this information to the test set composed of 5195 PIN, typed by 11 participants, while the other 5135 PINs composed by 11 participants were used to train the Base Attack. We first classified each PIN (input) of our test set as described above (i.e. 1038 was classified as Class 4 and 1004 was classified as Class 3). For each input we associated an array of all the PINs belonging to the same class and having the same different digits. We than performed a Base Attack filtering the output by removing all the PINs not included in the associated array.  Figure \ref {fig:thermal} shows the effectiveness of this algorithm comparing TA to Random Guess Thermal (RGT).

\subsection{Multple Attacks}\label{sec:combination}

\begin{figure}[h!]
	\centering
		\includegraphics[trim = 0.1in 2.5in 0.1in 2.5in, clip, scale=0.4]{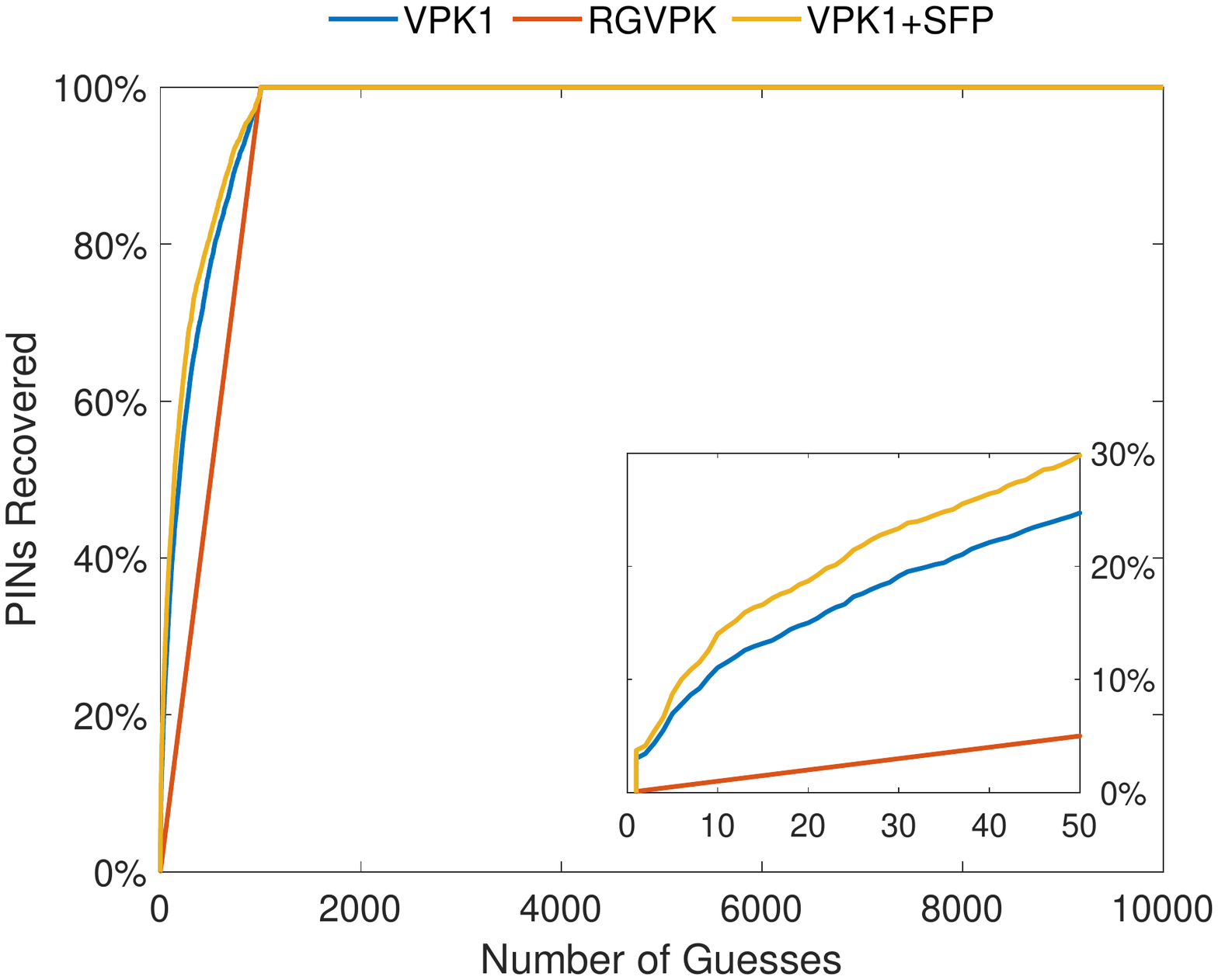}
	\caption{CDF showing the percentage of PINs recovered with Value-Position Knowledge of the first digit  (VPK1) , VPK1+ SFP (Single Finger PIN) and Random Guessing Value-Position Knowledge (RGVPK).}
	\label{fig:VPK_finger}
\end{figure}

\paragraph {Knowledge of one digit and its position in the code and of typing behavior.}\label{sec:VPK_finger}
We suppose that the adversary has come into possession of the value and the position of a digit in the code sequence (VPK) and that the PIN was a SFP (see Section \ref{sec:typing_behavior}). We investigated on real data how effective the combination of this knowledge is. We considered as testing set 3461 PINs typed by 11 participants containing only SFP as described in Section \ref{sec:typing_behavior}. We than performed to this testing set the VPK algorithm. Figure \ref {fig:VPK_finger} shows the effectiveness of this algorithm comparing the results to VPK and to RGVPK. Results show that the knowledge of typing behavior allows to improve considerably the performances for VPK \ref {fig:VPK_finger}.  In particular, within the first 5 attempts VPK+STP algorithm can guess 25\% more PINs respect to VPK. Moreover, VPK+STP shows a 11-fold improvement compared to RGVPK to guess PINs within 10 attempts.

\begin{figure}[h!]
	\centering
		\includegraphics[trim = 0.1in 2.5in 0.1in 2.5in, clip, scale=0.4]{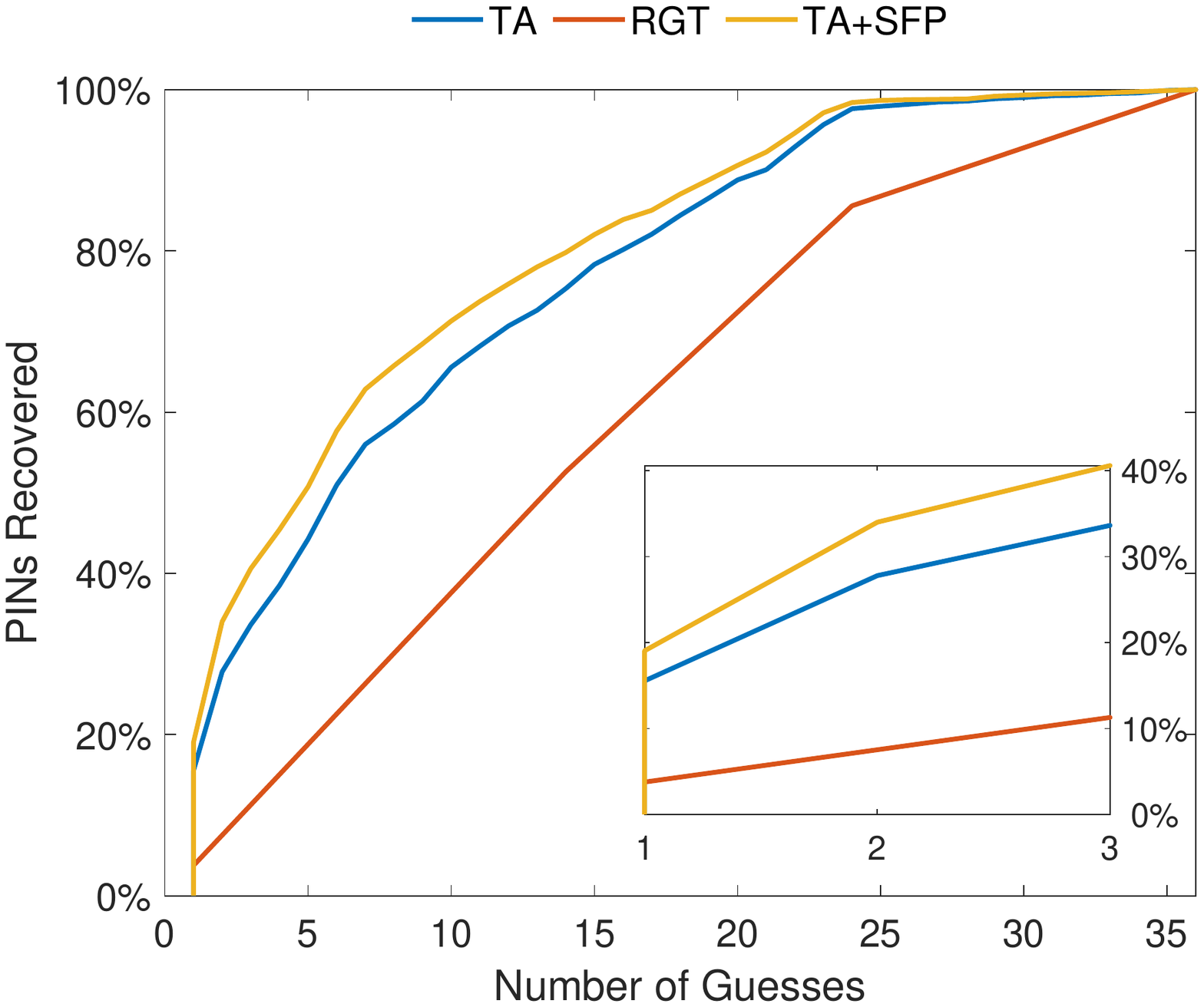}
	\caption{CDF showing the percentage of PINs recovered with Thermal Attack (TA),  Random Guess Thermal (RGT) and TA + SFP (Single Finger PIN).}
	\label{fig:thermal_finger}
\end{figure}

\paragraph{Knowledge of the values composing the PIN, but not their order and of typing behavior}\label{sec:thermal_finger}
We suppose that the adversary is aware of the keys pressed by the victim, but not of the temporal order in which they were pressed (Thermal Attack) and knows also that the typed PIN was a SFP (see Section \ref{sec:typing_behavior}). We investigated on real data how effective the combination of this knowledge is. We considered as testing set 3461 PINs typed by 11 participants containing only SFPs as described in Section \ref{sec:typing_behavior}. We than performed to this testing set the Thermal algorithm. Figure \ref {fig:thermal_finger} shows the effectiveness of this algorithm (Thermal+SFP) comparing the results to Thermal Attack and to Random Guess Thermal.

\begin{figure}[h!]
	\centering
		\includegraphics[trim = 0.1in 2.5in 0.1in 2.5in, clip, scale=0.4]{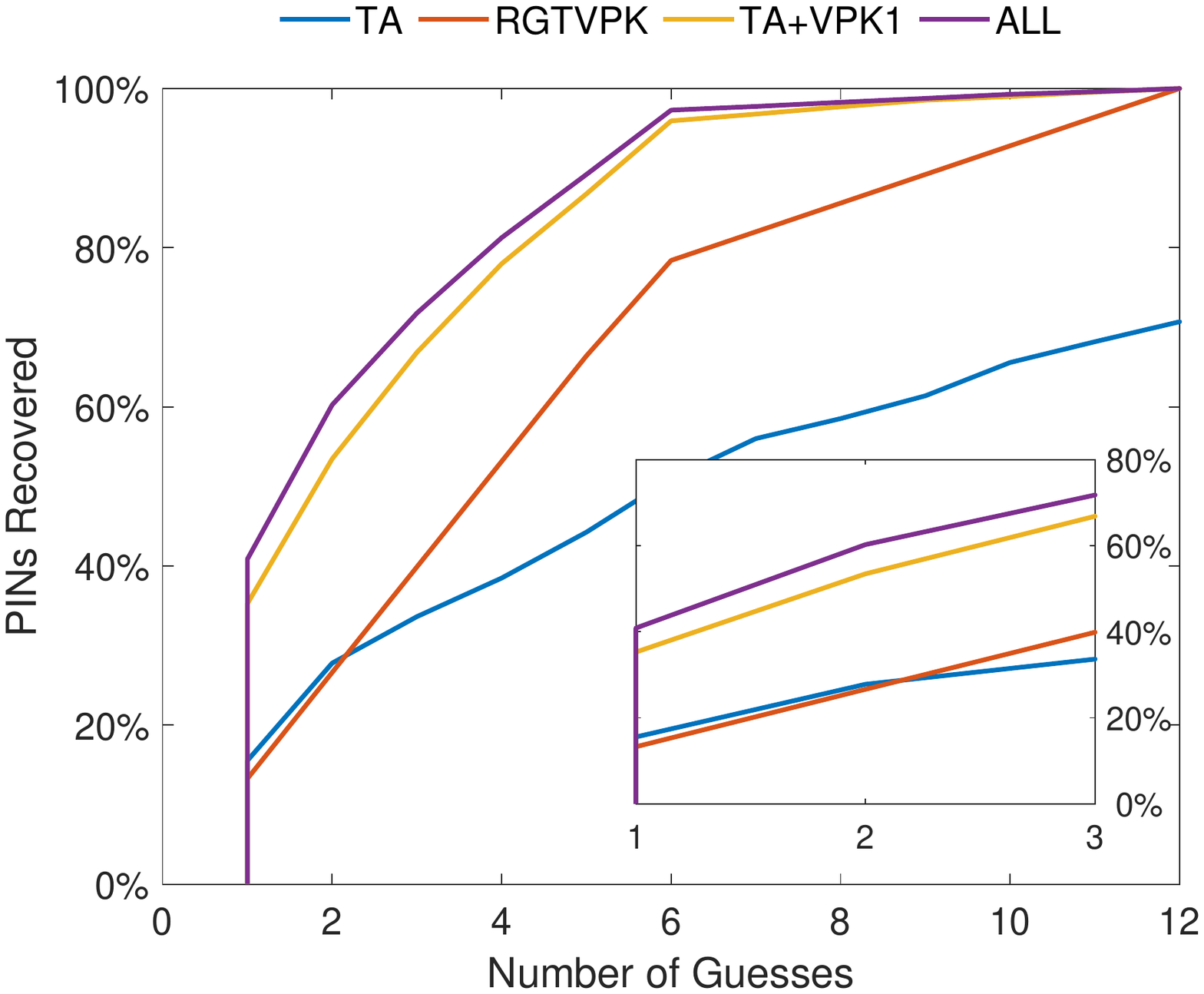}
	\caption{CDF showing the percentage of PINs recovered with Thermal Attack (TA), Random Guess Thermal + VPK (RGTVPK) ,TA+ VPK1 and TA + VPK1 + SFP (ALL).}
	\label{fig:ALL}
\end{figure}

\paragraph{Knowledge of the keys pressed when entering the code, but not the order except for one digit}\label{sec:thermal_VPK}
We suppose that the adversary is aware of the keys pressed by the victim, but not of the temporal order in which they were pressed (Thermal Attack) and knows also the position of one digit (VPK). In this scenario we performed the VPK and thermal algorithms in sequence. Figure \ref {fig:ALL} shows the effectiveness of this algorithm comparing Thermal+VPK to Thermal Attack and to Random Guess Thermal.

\paragraph{Knowledge of the keys pressed when entering the code, but not their order except for one digit and of typing behavior}\label{sec:ALL}
In this scenario it is assumed that the adversary is in possession of all the information: typing behavior (SFP), knowledge of one digit and its position in the code (VPK) and knowledge of the keys pressed when entering the code, but not the order (Thermal). We investigated on real data how effective the combination of this knowledge is. We considered as testing set 3461 PINs typed by 11 participants containing only SFPs as described in Section \ref{sec:typing_behavior}. We than performed to this testing set VPK and thermal algorithms in sequence.
Results show that the knowledge of all the conditions described in this paper and combined with the Base Attack, drastically increases the chance of guessing the PIN (see Figure \ref{fig:ALL}). In particular, within the first 3 attempts, the algorithm can identify about 72\% of PINs, offering an improvement of almost 2400 times compared to Random Guessing.

\section{PIN distribution}\label{sec:pin_dist}

The distribution of PIN codes is a factor that determines the security of systems that use this type of authentication. If 4-digit PINs are uniformly distribuited, the probability that any PIN is the correct one is 1 on 10,000 (0.01\%). In other words, an average of 5000 random attempts are required to guess the PIN.
In Section \ref {sec:pin_guessing} we showed that, depending on the information held by the adversary, it is possible to significantly reduce the number of attempts required to guess a PIN. The question we pose now is: does this information make a PIN more likely than another and therefore less secure?

As described in \cite{pilot} each input triplet of distances has a basket of PINs that can be associated. For instance the triplet 0, 0, 0 has 10 associated PINs  (i.e. 0000, 1111), while the triplet 1,1,1 has 216 associated PINs(i.e 0258, 4569). This heterogeneity makes some PINs more likely than others. Suppose that the adversary has succeeded in associating the exact Euclidean distance to each inter-keystroke timing, without errors. If the victim's real PIN is 5555, the triplet of input distances corresponds to 0, 0, 0. As we said, this triplet has 10 PINs associated and so the average attempts to guess the PIN are 5.5. Suppose instead that the real PIN is 8520, the triplet of input distances is 1,1,1. In this case the associated PINs are 216 and the average attempts to guess the PIN are 108.5. We can therefore affirm that the Base Attack affects the probability distribution for PINs. We verified this theory on real data. In this case the adversary, due to the overlap of inter-keystroke timing distributions \cite{pilot}, is not able to correctly estimate distances between keys, introducing a classification error. Our dataset consists of 44 PINs chosen to verify the correlation between inter-keystroke timing and Euclidean distance as described in \cite{silktv}. For each PIN we applied the Base Attack and calculated the number of attempts necessary to reach the 50\% probability to guess the PIN (P50). Table \ref{tab:P50} shows the 5 PINs with the lowest number of attempts for P50.

The knowledge of one digit of the code and of the typing behavior do not directly affect the distribution, although they significantly reduce the number of attempts to guess a PIN. The knowledge of one digit of the code allows to uniformly reduce the range of possible PINs from 10,000 to 1,000. For example, PIN 2200 and PIN 5555 have the same probability of 0.1\% to be guessed at the first attempt. However, in Section \ref{sec:VPK} we have seen that for our dataset, by applying the Base Attack to the knowledge of one digit of the code, the percentage of PINs guessed varies according to the position. In particular, knowing the first digit of the code (VPK1) or the fourth digit (VPK4) allows to significantly increase the percentage of PINs guessed in the first 10 attempts. This happens because VPK1 and VPK4 reduce classification error of the distances passed in input to the Base Attack more significantly than VPK2 or VPK3. However, this does not have a substantial reflection on the probability distribution of the PINs. As can be seen in Table \ref{tab:P50}, 4 on 5 PINs appear in the top 5 of both the BA and VPK1 algorithms, keeping the same positions. PINs 1038 (fourth in BA) and 3302 (fourth in VPK1), occupy the sixth position in VPK1 and in BA respectively.

The typing behavior is an information that can only be used in combination with the Base Attack (SFP). This knowledge increases the accuracy of the attack, but does not significantly affect the probability distribution of PINs. As reported in Table \ref{tab:P50}, 4 on 5 PINs appear in the top 5 of both the BA and SFP algorithms, keeping also the same positions. PIN 8556 (fifth in BA) and 2016 (third in SFP), occupy the ninth position in SFP and in BA respectively.

Another information that affects the probability distribution of PINs is the knowledge of the values composing the code. In Section \ref {sec:thermal} we described the 4 possible classes to which each PIN belongs. This shows how, depending on the class, the probability of each PIN changes. In particular, PINs belonging to Class 1 are the most probable, while those belonging to Class 3 are the least probable. 
Table \ref{tab:P50} reports the top 5 PINs ranking for our dataset, as the results of the combination of the knowledge of the values composing the PIN with the Base Attack (TA).

\tabulinesep = 2mm
\begin{table*}[!h]
\centering
\caption{Top 5 PIN for Base Attack (BA), Single Finger PIN (SFP), Value-Position Knowledge of the first digit (VPK1), Thermal Attack (TA) and number of attempts (Att.) necessary to reach the 50\% probability to guess the PIN}
\label{tab:P50}
\begin{tabu} to \textwidth {X[c]X[c]X[c]X[c]X[c]X[c]X[c]X[c]}
\toprule
\multicolumn{2}{c}{BA} & \multicolumn{2}{c}{SFP} & \multicolumn{2}{c}{VPK1} & \multicolumn{2}{c}{TA} \\
\midrule 
PIN      & Att    & PIN      & Att   & PIN      & Att   & PIN      & Att   \\
\midrule
2200     & 44          & 2200     & 6            & 2200      & 5            & 0489     & 1           \\
5555     & 87          & 1004     & 29           & 5555      & 5            & 5555     & 1           \\
1004     & 135         & 2016     & 85           & 1004      & 24           & 8556     & 1           \\
1038     & 213         & 1038     & 87           & 3302      & 30           & 9999      & 1          \\
8556     & 354         & 5555     & 95           & 8556      & 30           & 1004      & 2         \\
\bottomrule
\end{tabu}
\end{table*}

\section{Conclusion}\label{sec:conclusion}

In this paper we have shown that inter-keystroke timing retrieved from the feedback sound emitted by a PIN pad, can be effectively used to reduce the attempts to guess a PIN. We are able to correctly detect 98\% of feedback sound with an error of 0.13 +/- 6.66ms. We combined inter-keystroke timing with other information deduced from other plausible attacks in a real context. We demonstrated that inter-keystroke timing significantly improves performance for any attack: 1) if the adversary has come into possession of the value and the position of one digit from the PIN, inter-keystroke timing allows to reach a 14-fold improvement in performance within the first 5 attempts; 2) applying inter-keystroke timing to Thermal Attack allows to guess 15\% of the PINs at the first attempt, reaching a 4-fold improvement in performance.

We have demonstrated that the knowledge of typing behavior, combined with the inter-keystroke timing information, reduces number of attempts needed to guess the PIN. In particular, if a PIN is typed always with the same finger (SFP) the adversary is able to guess it with a lower number of attempts compared to a PIN entered with multiple fingers or two hands (OP). Our results show that knowledge of SFP leads to a 34-fold improvement compared to random guessing, resulting in PINs guessed within 5 attempts. Moreover, among the users that participated in our study, 73\% prefer to type the PIN using only one finger, while the remaining 27\% type using more fingers than a hand or two hands.

We have shown how the combination of multiple attacks can dramatically reduce attempts to guess the PIN. In particular, we were able to guess 72\% of the PINs within the first 3 attempts, and almost 90\% of the PINs within 5 attempts. Our results highlight a real threat to PIN authentication systems. The feasibility of the attack and its immediate applicability in real scenarios poses a considerable security threat for ATMs, PoS-s, and similar devices.

Finally, we have shown how the knowledge of inter-keystroke timing significantly improves the guessing probability of certain subsets of PINs. This poses a serious problem in the security of these classes of PINs, and as a result our work suggests that secre PIN generation must take this into account.

\bibliography{bibliography}

\begin{thebibliography}{10}
\providecommand{\url}[1]{\texttt{#1}}
\providecommand{\urlprefix}{URL }
\providecommand{\doi}[1]{https://doi.org/#1}

\bibitem{thermalCamera}
Flir one thermal imaging camera for ios (gen 3).
  \url{https://www.amazon.com/FLIR-ONE-Thermal-Imaging-Camera/dp/B071Z63RSL},
  accessed: 2019-04-23

\bibitem{abdelrahman2017stay}
Abdelrahman, Y., Khamis, M., Schneegass, S., Alt, F.: Stay cool! understanding
  thermal attacks on mobile-based user authentication. In: Proceedings of the
  2017 CHI Conference on Human Factors in Computing Systems. pp. 3751--3763.
  ACM (2017)

\bibitem{anderson1952asymptotic}
Anderson, T.W., Darling, D.A., et~al.: Asymptotic theory of certain" goodness
  of fit" criteria based on stochastic processes. The annals of mathematical
  statistics  \textbf{23}(2),  193--212 (1952)

\bibitem{asonov2004keyboard}
Asonov, D., Agrawal, R.: Keyboard acoustic emanations. In: IEEE S\&P (2004)

\bibitem{pilot}
Balagani, K., Cardaioli, M., Conti, M., Gasti, P., Georgiev, M., Gurtler, T.,
  Lain, D., Miller, C., Molas, K., Samarin, N., et~al.: Pilot: Password and pin
  information leakage from obfuscated typing videos. arXiv preprint
  arXiv:1904.00188  (2019)

\bibitem{silktv}
Balagani, K.S., Conti, M., Gasti, P., Georgiev, M., Gurtler, T., Lain, D.,
  Miller, C., Molas, K., Samarin, N., Saraci, E., et~al.: Silk-tv: Secret
  information leakage from keystroke timing videos. In: European Symposium on
  Research in Computer Security. pp. 263--280. Springer (2018)

\bibitem{batiz2011development}
B{\'a}tiz-Lazo, B., Reid, R.: The development of cash-dispensing technology in
  the uk. IEEE Annals of the History of Computing  \textbf{33}(3),  32--45
  (2011)

\bibitem{berger2006dictionary}
Berger, Y., Wool, A., Yeredor, A.: Dictionary attacks using keyboard acoustic
  emanations. In: Proceedings of the 13th ACM conference on Computer and
  communications security. pp. 245--254. ACM (2006)

\bibitem{bonneau2012birthday}
Bonneau, J., Preibusch, S., Anderson, R.: A birthday present every eleven
  wallets? the security of customer-chosen banking pins. In: International
  Conference on Financial Cryptography and Data Security. pp. 25--40. Springer
  (2012)

\bibitem{butterworth1930theory}
Butterworth, S.: On the theory of filter amplifiers. Wireless Engineer
  \textbf{7}(6),  536--541 (1930)

\bibitem{halevi2012closer}
Halevi, T., Saxena, N.: A closer look at keyboard acoustic emanations: random
  passwords, typing styles and decoding techniques. In: Proceedings of the 7th
  ACM Symposium on Information, Computer and Communications Security. pp.
  89--90. ACM (2012)

\bibitem{iso9564}
ISO: Financial services -- personal identification number (pin) management and
  security -- part 1: Basic principles and requirements for pins in card-based
  systems (2017), \url{https://www.iso.org/standard/68669.html}

\bibitem{kaczmarek2018thermanator}
Kaczmarek, T., Ozturk, E., Tsudik, G.: Thermanator: Thermal residue-based post
  factum attacks on keyboard password entry. arXiv preprint arXiv:1806.10189
  (2018)

\bibitem{kumar2007reducing}
Kumar, M., Garfinkel, T., Boneh, D., Winograd, T.: Reducing shoulder-surfing by
  using gaze-based password entry. In: Proceedings of the 3rd symposium on
  Usable privacy and security. pp. 13--19. ACM (2007)

\bibitem{kwon2015analysis}
Kwon, T., Hong, J.: Analysis and improvement of a pin-entry method resilient to
  shoulder-surfing and recording attacks. Ieee transactions on information
  forensics and security  \textbf{10}(2),  278--292 (2015)

\bibitem{liu2015snooping}
Liu, J., Wang, Y., Kar, G., Chen, Y., Yang, J., Gruteser, M.: Snooping
  keystrokes with mm-level audio ranging on a single phone. In: Proceedings of
  the 21st Annual International Conference on Mobile Computing and Networking.
  pp. 142--154. ACM (2015)

\bibitem{marquardt2011sp}
Marquardt, P., Verma, A., Carter, H., Traynor, P.: (sp) iphone: Decoding
  vibrations from nearby keyboards using mobile phone accelerometers. In:
  Proceedings of the 18th ACM conference on Computer and communications
  security. pp. 551--562. ACM (2011)

\bibitem{mowery2011heat}
Mowery, K., Meiklejohn, S., Savage, S.: Heat of the moment: Characterizing the
  efficacy of thermal camera-based attacks. In: Proceedings of the 5th USENIX
  conference on Offensive technologies. pp.~6--6. USENIX Association (2011)

\bibitem{roth2004pin}
Roth, V., Richter, K., Freidinger, R.: A pin-entry method resilient against
  shoulder surfing. In: Proceedings of the 11th ACM conference on Computer and
  communications security. pp. 236--245. ACM (2004)

\bibitem{sarkisyan2015wristsnoop}
Sarkisyan, A., Debbiny, R., Nahapetian, A.: Wristsnoop: Smartphone pins
  prediction using smartwatch motion sensors. In: 2015 IEEE international
  workshop on information forensics and security (WIFS). pp.~1--6. IEEE (2015)

\bibitem{song2001timing}
Song, D.X., Wagner, D., Tian, X.: Timing analysis of keystrokes and timing
  attacks on ssh. In: USENIX Security Symposium (2001)

\bibitem{vuagnoux2009compromising}
Vuagnoux, M., Pasini, S.: Compromising electromagnetic emanations of wired and
  wireless keyboards. In: USENIX security symposium. pp. 1--16 (2009)

\bibitem{wang2016friend}
Wang, C., Guo, X., Wang, Y., Chen, Y., Liu, B.: Friend or foe?: Your wearable
  devices reveal your personal pin. In: Proceedings of the 11th ACM on Asia
  Conference on Computer and Communications Security. pp. 189--200. ACM (2016)

\bibitem{wang2017understanding}
Wang, D., Gu, Q., Huang, X., Wang, P.: Understanding human-chosen pins:
  characteristics, distribution and security. In: Proceedings of the 2017 ACM
  on Asia Conference on Computer and Communications Security. pp. 372--385. ACM
  (2017)

\bibitem{wang2014ubiquitous}
Wang, J., Zhao, K., Zhang, X., Peng, C.: Ubiquitous keyboard for small mobile
  devices: harnessing multipath fading for fine-grained keystroke localization.
  In: Proceedings of the 12th annual international conference on Mobile
  systems, applications, and services. pp. 14--27. ACM (2014)

\bibitem{zalewski2005cracking}
Zalewski, M.: Cracking safes with thermal imaging. ser. http://lcamtuf.
  coredump. cx/tsafe  (2005)

\bibitem{zhu2014}
Zhu, T., Ma, Q., Zhang, S., Liu, Y.: Context-free attacks using keyboard
  acoustic emanations. In: ACM CCS (2014)

\end{thebibliography}

\end{document}